\documentclass{mn2e}
\usepackage{times}

\input{psfig.sty}

\begin{document}

\title[Neutron stars' jets]
{Jets in neutron star X-ray binaries: a comparison with black holes}

\author[S. Migliari \& R.P. Fender]
{S. Migliari$^{1,2}$\thanks{migliari@ucsd.edu}, R. P. Fender$^{2,3}$\thanks{rpf@phys.soton.ac.uk}\\ 
\\
$^1$ Center for Astrophysics and Space Sciences, University of California San Diego, 9500 Gilman Dr., La Jolla, CA 92093-0424\\
$^2$ Astronomical Institute `Anton Pannekoek', University of Amsterdam, and Center for High Energy Astrophysics, Kruislaan 403, \\  
1098 SJ, Amsterdam, The Netherlands.\\
$^3$ School of Physics and Astronomy, University of Southampton
Hampshire SO17 1BJ, United Kingdom 
}

\maketitle

\begin{abstract}
We present a comprehensive study of the relation between radio and
X-ray emission in neutron star X-ray binaries, use this to infer the
general properties of the disc--jet coupling in such systems, and
compare the results quantitatively with those already established for
black hole systems. There are clear qualitative similarities between
the two classes of object: hard states below about 1\% of the
Eddington luminosity produce steady jets, while transient jets are
associated with outbursting and variable sources at the highest
luminosities. However, there are important quantitative differences:
the neutron stars are less radio-loud for a given X-ray luminosity
(regardless of mass corrections), and they do not appear to show the
strong suppression of radio emission in steady soft states which we
observe in black hole systems. Furthermore, in the hard states the
correlation between radio and X-ray luminosities of the neutron star
systems is steeper than the relation observed in black holes by about
a factor of two. This result strongly suggests that the X-ray emission
in the black hole systems is radiatively inefficient, with an
approximate relation of the form $L_X \propto \dot{m}^2$, consistent
with both advection-dominated models and jet-dominated scenario. On
the contrary the jet power in both classes of object scales linearly
with accretion rate. This constitutes some of the first observational
evidence for the radiatively inefficient scaling of X-ray luminosity
with accretion rate in accreting black hole systems. Moreover, based
on simultaneous radio/X-ray observations of Z-type neutron stars (the
brightest of our galaxy, always near or at the Eddington accretion
rate), we draw a model that can describe the disc-jet coupling in such
sources, finding a possible association between a particular X-ray
state transition (horizontal brach-to-normal branch) and the emission
of transient jets.  \\
\end{abstract}


\section{Introduction}

Multiwavelength studies of X-ray binaries (XRBs), especially in the
past decade, have shown that a significant fraction of the dissipated
accretion power may be released in form of radiatively inefficient
collimated outflows, or jets. In general, relativistic jets are very
common features associated to accretion onto relativistic compact
objects on all mass scales, from neutron stars (NSs) and stellar-mass
black holes (BHs) in XRB systems to supermassive BHs in active
galactic nuclei (AGN), and thought to be at the origin of gamma-ray
bursts (GRBs), the most powerful transient phenomena in the
universe. The advantage of studying relativistic jets in XRBs is
mainly due to the fact that the accretion process varies on much
faster (humanly-accessible) timescales than in AGN, allowing us to
observe and follow significant evolution of the systems, and to
investigate the link between the jet production and the different
accretion regimes. At present, most of our knowledge about jets in
XRBs has come from studies of black hole candidates. This is mainly
due to the fact that in general, exceptions are the so-called Z-type
NSs, BH XRBs are more radio loud than NSs, hence easier to detect.

\subsection{Black hole X-ray binaries}

A non-linear correlation has been found, linking the radio to the X-ray
luminosities in BH XRB systems over more than three orders of magnitude in
X-rays, when the BHs are in hard state. This relation takes the form
$L_{R}\propto L_{X}^b$ where $L_{R}$ and $L_{X}$ are the radio and X-ray
luminosities, and $b \sim 0.7$ (Corbel et al. 2003; Gallo, Fender \& Pooley
2003). In the hard state (i.e. below a few percent the Eddington luminosity),
the radio emission is observed to be optically thick, with a flat or slightly
inverted spectrum, and although a jet has been spatially resolved only in two
sources so far (Cyg~X-1: Stirling et al. 2001; GRS~1915+105: Dhawan, Mirabel \&
Rodr\'\i guez 2000), indirect evidence indicates that this is the signature of
a continuously replenished steady jet, the so-called `compact jet' (see Fender
2005 for a review).

In BH XRBs (but maybe also in AGN, see Maccarone, Gallo \& Fender 2003) there
is evidence for a quenching of radio emission when the source is steadily in
the soft state, probably due to a physical suppression of the jet (Fender et
al. 1999; Gallo et al. 2003). The rapid X-ray transition from hard to soft
states [i.e. very-high (VHS) or steep power-law state] is associated with
radio flares which show optically thin spectra. These radio flares are the
signatures of powerful ejection events, spatially resolved as large-scale
(from tens to thousands of milliarcsec) extended jets (e.g. Mirabel \&
Rodriguez 1994; Hjellming \& Rupen 1995; Fender et al. 1999; Corbel et
al. 2001; Gallo et al. 2004). A unified semiquantitative model for the
disc-jet coupling in BH XRBs, covering both steady and transient jets, has
been presented by Fender, Belloni \& Gallo (2004)

Extending the correlation found for BH XRBs in the hard state also to
supermassive BHs, and with the addition of the mass parameter, there is
evidence for a `fundamental plane of BH activity' in which a single 3D
power-law function can fit all the BH data (XRBs and AGN) for a given X-ray
luminosity, radio luminosity and mass of the compact object.  This plane takes
the approximate form $L_{R}\propto L_{X}^{0.6} M^{0.8}$ where $M$ is the mass
of the compact object (Merloni, Heinz \& Di Matteo 2003; Falcke, K\"ording \&
Markoff 2004). The existence of this relation connecting BH XRBs and AGN
points towards the same physical processes as drivers of the disc-jet
coupling, regardless of the mass of the BH involved. The radio:X-ray
luminosity power-law correlation previously found by studying only BH XRBs
has, within uncertainties, the same slope found in the correlation of the
`fundamental plane' with BH XRBs and AGN. Clearly the study of XRBs can be
fundamental for our understanding of the physical properties of discs and jets
in compact objects in general, including supermassive black holes in the
centres of distant galaxies.

\subsection{Neutron star X-ray binaries}

\begin{figure}
\centerline{\psfig{file=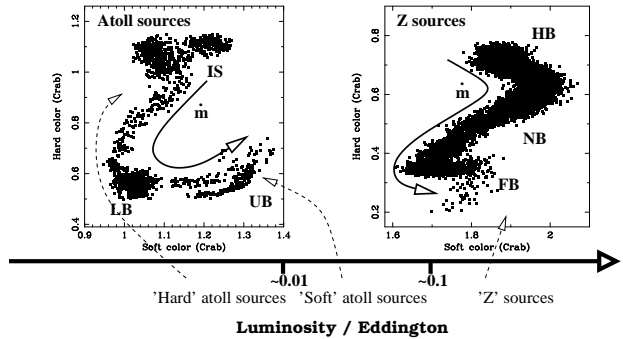,width=9cm,angle=270}}
\vspace*{-1cm}
\caption{A summary of the simplified description of the relation of
accretion / luminosity to patterns of X-ray behaviour, for low magnetic field
neutron stars, as adopted in this paper.  For both types of sources the
expected direction of increasing mass accretion rate, $\dot{m}$, is
indicated. For the atoll sources, the acronyms are: IS = island state, LB =
lower banana; UB = upper banana.  For the Z sources the acronyms are: HB =
horizontal branch, NB = normal branch, FB = flaring branch. CDs are from van
Straaten et al. (2003) and Jonker et al. (2000).}\label{ns-types}
\end{figure}

Low-magnetic field NS XRBs have been classified, based on their X-ray spectral
and timing properties, in two main distinct classes whose names derive from
the shape they trace in the X-ray colour-colour diagram (CD): Z-type and
atoll-type NSs (see Hasinger \& van der Klis 1989). The 'broad' definition of
atoll sources as having the apparent characteristics of low magnetic field
neutron stars accreting at relatively low rates (compared to the Z sources)
rather oversimplifies the more complex classifications of e.g. 'burster',
'dipper', etc., but is appropriate for the discussion in this
paper. Fig.~\ref{ns-types} summarises the simplified classifications adopted
in this paper (see van der Klis 2005 for a more detailed classification).

The galactic Z-type NS XRBs represent a class of six low-mass XRBs (possibly
seven if we include Cir~X-1 which may be considered as a `peculiar' Z source;
Shirey, Bradt \& Levine 1999) accreting near or at the Eddington rate, and are
the most luminous NS XRBs in our Galaxy. The name of this class of sources
derives from the typical `Z' track traced by their CD
(Fig.~\ref{ns-types}). The three branches which form the Z-shaped CD are
called Horizontal (HB), Normal (NB) and Flaring (FB), top-left to
bottom-right, and define three distinct spectral states of the systems. Z
sources are rapidly variable in X-rays and can trace the whole CD, transiting
in the different states, in hours to days. This variability is thought to be
physically related to changes in the mass accretion rate, which should
increase along the Z-track from HB to FB (Hasinger \& van der Klis 1989). In
the radio band, we also observe large and rapid variability, optically thick
and optically thin emission.  All the Z-type NS sources have been detected in
radio.  Looking in detail at the radio behaviour of Z sources as a function of
X-rays, Penninx et al. (1988) first found in GX~17+2 a qualitative relation
between disc and jet properties: the radio emission varies as a function of
the position in the X-ray CD, decreasing with increasing (inferred) mass
accretion rate from HB (strongest radio emission) to FB (weakest radio
emission). A behaviour consistent with GX~17+2 has been found also in Cyg~X-2
(Hjellming et al 1990a) and Sco~X-1 (Hjellming et al 1990b). An exception is
GX~5-1, which showed a low and steady radio flux when the source was in the
HB, then increasing when in the NB (Tan et al. 1992).  Extended radio jets
have been spatially resolved for two Z sources: Sco~X-1 (Fomalont et
al. 2001a) and Cir~X-1 (Fender et al. 1998). In this two sources there is also
evidence for an association between radio flares and powerful
(ultra-relativistic) ejections from the system (Fomalont et al. 2001b; Fender
et al. 2004).

Atoll-type NS XRBs share many X-ray spectral and timing properties with BH
XRBs and show two distinct (hard and soft) X-ray states, defined by the
position in the CD, that can be directly compared to the hard and soft state
of BH XRBs: the hardest X-ray state is called `island' and the softest
`banana' (Fig.~\ref{ns-types}).  Although atolls represent the largest class
of known X-ray binaries, only a few are detected in the radio band because of
their lower radio luminosity ($\sim 30$ times less 'radio loud' than BH and
Z-type NS XRBs: Fender \& Kuulkers 2001; Migliari et al. 2003; Muno et
al. 2005; this paper).  To date, five atoll sources have been detected in the
radio band during simultaneous radio/X-ray observations: 4U~1728-34,
4U~1820-30, Ser~X-1, Aql~X-1 and MXB~1730-335 (Migliari et al. 2003, 2004;
Rupen et al. 2004; Rutledge et al. 1998; Moore et al. 2000). In particular,
4U~1728-34, which is to date the only atoll source detected in radio when
steady in its hard state, shows a positive correlation between radio and X-ray
fluxes, similar to what observed in BHCs (Migliari et al. 2003). Homan et
al. (2004) have also investigated the `peculiar' atoll source GX~13+1 which is
persistently at a very high X-ray luminosity of a few tens per cent Eddington,
showing that its radio behaviour is much more similar to Z sources

Two accreting ms X-ray pulsars, SAX J1808.4-3658 (e.g. Gaensler, Stappers
\& Getts 1999) and IGR~J00291+5934 (Pooley 2004), have
shown transient radio emission related to X-ray outbursts. These flares may be
signatures of transient relativistic outflows from the system as observed in
BH XRBs and Z sources. In the context of the general picture of low magnetic
field neutron stars, we assume (initially at least) that these systems behave
like atoll sources, although Chakrabarty (2005) has suggested they may have
systematically higher magnetic fields.

None of the high-magnetic field X-ray pulsars has ever been convincingly
detected as a synchrotron radio source (e.g. Fender \& Hendry 2000 and
references therein). This has been explained by the fact that surface
high-magnetic field, which probably disrupts the inner regions of the
accretion disc around the NS (e.g. White, Nagase \& Parmar 1995; Bildsen et
al. 1997) or which may strongly interact with the jet magnetic fields coupled
with the inner regions of the disc (Migliari, Fender \& van der Klis 2005),
results in suppressed jet formation (Fender \& Hendry 2000).

\section{The sample}

\begin{table*}
\centering
\caption{Name of the source of our sample, X-ray state (LB=lower banana;
IS=island; MB=middle-banana; H-OUTB=hard outburst; S-OUTB=soft outburst;
ASM/PCA mean=the mean over more than one X-ray state), X-ray flux in
the range 2-10~keV, radio flux density at 8.5~GHz, distance to the source in
kpc; references} 
\begin{tabular}{l l l l l l}
\hline
Source & X-ray state & F$_{2-10}$ ($\times10^{-9}$ erg~cm$^{-2}$~s$^{-1}$)&
F$_{8.5}$ (mJy) & D (kpc) & Ref.\\
\hline
\multicolumn{4}{c}{Atoll-type NSs} \\
\hline
4U 1728-34 &LB&$1.03\pm0.05$&$0.5\pm0.08$&4.6 &M03,G03\\           
           &IS&$2.25\pm0.15$& $0.6\pm0.2$& &\\
           &LB& $1.54\pm0.10$& $0.33\pm0.15$& &\\
           &IS&$1.81\pm0.11$ & $0.62\pm0.1$& &\\
           &IS&$2.42\pm0.16$ & $0.11\pm0.02$& &\\
           &IS&$0.60\pm0.07$ & $0.09\pm0.02$& &\\
           &IS&$0.61\pm0.06$ & $0.11\pm0.02$& &\\
           &IS&$0.62\pm0.06$ & $0.15\pm0.02$& &\\
           &IS&$0.69\pm0.12$ & $0.16\pm0.02$& &\\
           &IS&$0.70\pm0.05$ & $0.09\pm0.02$& &\\
4U 1820-30 &MB &$8.7$ &$0.10\pm0.02$ & 7.6&M04,H00\\
Ser X-1    &MB &$4.4$ &$0.08\pm0.02$ & 12.7&M04,JN04\\
Aql X-1& H-OUTB &$0.79$&$0.210\pm0.050$&5.2 &RMD04,JN04\\
       & H-OUTB & $1.00$& $0.214\pm0.035$& &\\
4U 1608-52& IS?& $0.93$& $<0.19$&3.3 &MF05,JN04\\
4U 0614-09& IS& $0.78$& $<0.09$& $<3$ &MF05,B92\\
MXB 1730-335 &S-OUTB&$2.92$ & $0.370\pm0.030$ &8.8&R98,M00,K03\\
             &S-OUTB&$3.06$ & $0.290\pm0.030$ & &\\
             &S-OUTB&$5.34$ & $0.330\pm0.050$ & &\\
\hline
\multicolumn{4}{c}{Low-magnetic field accreting ms X-ray pulsar} \\
\hline
SAX J1808.4-3658& S-OUTB & $0.14$& $0.8\pm0.18$&2.5&GSG99,Z01,R02,R05\\
                & S-OUTB & $1.82$& $0.44\pm0.06$&2.5& \\
                & S-OUTB & $0.56$& $0.44\pm0.07$&2.5&\\
IGR J00291+5934& S-OUTB &$0.62$ & $1.5\pm0.3$& $<3$?&G05,P04\\
\hline
\multicolumn{4}{c}{Z-type NSs} \\
\hline
Sco X-1 &ASM mean & $253.80$& $10\pm3$&$2.8$&FH00,P89,C76,BFG97 \\
GX 17+2 &ASM mean & $12.90$& $1.0\pm0.3$&$14$&FH00,P89,P88,JN04 \\
GX 349+2& ASM mean& $14.39$& $0.6\pm0.3$&$5$& FH00,CP91,CS97\\
Cyg X-2 &ASM mean & $10.75$& $0.6\pm0.2$&$13.3$ & FH00,P89,H90,CCH79,JN04\\
GX 5-1  &ASM mean & $20.36$& $1.3\pm0.3$&$9.2$ & FH00,P89\\
GX 340+0&ASM mean & $8.54$& $0.6\pm0.3$&$11$ & FH00,P93\\
GX 13+1 &PCA mean & $18$& $1.8\pm0.3$&$7$ & FH00,H04,JN04\\
\hline
\multicolumn{4}{c}{High-magnetic field accreting X-ray pulsars} \\
\hline
X Per  & ?&$0.46$ &$<0.08$ &1 &MF05,D01\\
4U 2206+54 & ?& $0.26$ &$<0.039$& 3&B05,NR01\\
\hline

\end{tabular}
\flushleft
{\em Refs:} M03=Migliari et al. 2003; G03=Galloway et al. 2003; M04=Migliari
et al. 2004; H00=Heasley et al. 2000; JN04=Jonker \& Nelemans 2004;
RMD04=Rupen, Mioduszewski \& Dhawan 2004; MF05=Migliari \& Fender 2005, in
prep.; B92=Brandt et al. 1992; R98=Rutledge et al. 1998; M00=Moore et
al. 2000; K03=Kuulkers et al. 2003 and references therein; GSG99=Gaensler,
Stappers \& Getts 1999; Z01=in 't Zand et al. 2001; R02=Rupen et al. 2002;
R05=Rupen et al. 2005; G05=Galloway et al. 2005; P04=Pooley 2004; FH00=Fender
\& Hendry 2000; P89=Penninx 1989; C76=Crampton et al. 1976; BFG97=Bradshaw,
Fomalont \& Geldzahler 1997; P88=Penninx et al. 1988; CP91=Cooke \& Ponman
1991; CS97=Christian \& Swank 1997; H90=Hjellming et al. 1990; CCH79=Cowley,
Crampton \& Hutchings 1979; P93=Penninx et al. 1993; H04=Homan et al. 2004;
D01= Delgado-Mart\'\i\ et al. 2001; B05=Blay et al. 2005; NR01=Negueruela \&
Reig 2001.
\end{table*}

In Table~1 we list the names, X-ray states, fluxes and estimated distances of
all the NS XRBs in our sample. 

\subsection{New radio observations of NS XRBs}

The atoll-type NS XRB 4U~1608-52 (located at a distance of $\sim3.3$~kpc;
Jonker \& Nelemans 2004) has been observed in 2004, on March 31 and April 03
for a total of $\sim12$~hr with the Australian Telescope Compact Array (ATCA)
at 8.5~GHz. During the observation ATCA was in configuration 1.5A. We have
analysed the data with the package MIRIAD (Sault, Teuben \& Wright 1995),
using PKS 1934-638 as flux calibrator and 1646-50 as phase calibrator. At the
best (X-ray) coordinates, RA $16^{h}12^{m}43^{s}.0$ Dec
$-52^{\circ}25^{\prime}23\arcsec$, we did not detect the radio counterpart
with a $3\sigma$ radio flux density upper limit of $F<0.19$~mJy.  The source
was in a quiescent (hard) X-ray state with a mean (two-day average) count rate
of $3.21\pm0.15$ in the All-Sky Monitor (ASM; 2-10~keV) on board the Rossi
X-ray Timing Explorer (RXTE). The X-ray flux reported in Table~1 have been
estimated converting the ASM count rates to Crab based on Levine et al. (1996:
1~Crab=75~c/s) and then to 2-10~keV flux.

The atoll-type NS XRB 4U~0614-09 (which is at a distance of less than 3~kpc;
Brandt et al. 1992) has been observed on 2001 April 24 for 12~hr with the
Westerbork Synthesis Radio Telescope (WSRT) at 5~GHz.
We have analysed the data with the package MIRIAD, using 3C286 as flux
calibrator. At the best (optical) coordinates, RA 06 17 07.3 Dec +09 08 13, we
did not detect the radio counterpart with a $3\sigma$ radio flux density upper
limit of $F<0.09$~mJy.  The source was in a quiescent (hard) X-ray state with
a one-day averaged RXTE/ASM count rate of $2.71\pm0.81$~c/s. 

The XRB pulsar X Per (the nearest known, at a distance of only
$0.7\pm0.3$~kpc) has been observed on 2004 November 25 for 12~hr with the
Westerbork Synthesis Radio Telescope (WSRT) at 1.4~GHz.
We have analysed the data with the package MIRIAD, using 3C286 as flux
calibrator. At the best (optical) coordinates, RA $03^{h}55^{m}23^{s}.08$ Dec
$+31^{\circ}02^{\prime}45\arcsec.0$, we did not detect the radio counterpart
with a $3\sigma$ radio flux density upper limit of $F<0.08$~mJy. The
simultaneous one-day averaged RXTE/ASM count rate was $1.61\pm0.34$~c/s.
 
No radio counterparts of the X-ray sources 4U~1608-52, 4U~0614-09 and X~Per
have ever been detected and these are the most stringent upper limits to date.

\subsection{Other atoll-type NS XRBs}
\noindent

{\em 4U 1728-34}: based on two distinct epochs of observations, we found
correlations between radio flux density and both X-ray flux and X-ray timing
features (Migliari et al. 2003). One data point -- that with the highest X-ray
flux -- lay well off the correlation, possibly indicating `quenching' of the
jet as observed in black holes (but see discussion in Migliari et al. 2003).
4U~1728-34 has been detected in the radio band, both when the source was
steadily in its hard state (island), and when it was repeatedly transiting
between the island state and a softer state (lower banana). The strongest and
most variable radio emission seems to be related to the X-ray state
transitions. These radio detections also allowed us to quantify the difference
in radio power between BH and NS XRBs when steadily in their hard state: NSs
are a factor of $\sim30$ less `radio loud' than BHs (at a soft X-ray
luminosity of $\sim 10^{36}$ erg s$^{-1}$). Setting the upper limit on the
brightness temperature of the radio emitting region to $10^{12}$~K (above
which, for steady states, inverse Compton losses will rapidly cool the
plasma), we can estimate a lower limit on the size R of the emitting region of
R$> 1.4\times 10^{11}$~cm, likely to be larger than the binary stars
separation [typical low-luminosity ($10^{36}-10^{37}$~erg/s) low-mass XRBs
have stars separations smaller than $\sim3\times10^{11}$~cm; White et
al. 1995], and thus unbound to the system. The dual-band radio spectra of
these observations are consistent with a steady jet emission as observed in BH
XRBs in the hard state (even if errors on the detections cannot definitely
rule out an optically thin spectrum). In Table~1 we show X-ray and radio
fluxes reported in Migliari et al. (2003) and the mean of the estimated
distances in Galloway et al. (2003).

{\em 4U 1820-30 and Ser X-1}: we detected the radio counterparts of 4U~1820-30
and Ser~X-1 when the sources were steadily in their soft X-ray state
(lower-banana; Migliari et al. 2004). From the measured radio flux densities
we estimate the size of the radio emitting regions to be R$>7\times10^{10}$~cm
for 4U 1820-30, larger than the star binary system separation
($\sim1.3\times10^{10}$~cm; e.g. Arons \& King 1993), and R$>10^{11}$~cm for
Ser X-1, also likely to be larger than the binary stars separation (White et
al. 1995). In Table~1 we show the X-ray and radio fluxes reported in Migliari
et al. (2004) and the estimated distances from Jonker \& Nelemans (2004).
 
{\em Aql X-1}: Rupen et al. (2004) reported a radio transient emissions in
Aql~X-1 associated with an X-ray, optical and IR outburst of the source.  Note
that during the X-ray outburst the source never entered the soft state (see
also Reig, van Straaten \& van der Klis 2004). Hereafter, we will refer to
such an outburst as a `hard' X-ray outburst, while we will call `soft'
outbursts, X-ray outbursts during which the source enters the soft state. On
2004 May 26 two quasi-simultaneous observations at 8.5~GHz and 5~GHz, give a
spectral index $\alpha=+0.4\pm0.8$ (where $S_{\nu}\propto\nu^{\alpha}$ and
$S_{\nu}$ is the radio flux density at a frequency $\nu$) which seems to
suggest optically-thick emission typical of BH XRBs in their hard state,
although uncertainties on the estimated flux densities cannot rule out the
possibility of an optically-thin emission (usually with $\alpha\sim-0.6$). In
Table~1, we report the mean of the estimated distances in Jonker \& Nelemans
(2004), the flux densities of the radio detections reported in Rupen et
al. (2003), and the X-ray flux calculated from simultaneous RXTE/ASM
(2-10~keV) daily-averaged observations.

{\em MXB 1730-335 (the Rapid Burster)}: Rutledge et al. (1998) and Moore et
al. (2000) reported simultaneous observations with RXTE/ASM and the Very Large
Array (VLA) at 5~GHz and 8.5~GHz that revealed a transient radio emission
correlated with the X-ray flux, during a (soft) X-ray outburst. The
dual-frequency radio observations indicate a flat or slightly inverted
spectral index. In Table~1 we show the X-ray and radio fluxes reported in
Reynolds et al. (1998) and the distance in Kuulkers et al. (2003) and
references therein.

\subsection{Low-magnetic field accreting ms X-ray pulsars}
\noindent

{\em SAX J1808.4-3658}: Gaensler et al. (1999) reported a radio detection of
SAX J1808.4-3658 during the (soft) X-ray outburst in 1998. The size of the
radio emission region can be constrained to be R$>3.6\times10^{10}$~cm, larger
than the binary stars separation (Chakrabarty \& Morgan 1998). In Table~1 we
show radio flux densities reported in Gaensler et al. (1998) and the estimated
distance from in't Zand et al. (2001).  Rupen et al. (2002 and 2005) reported
radio detections at 8.5~GHz during the X-ray outbursts on 2002 October 16 and
on 2005 June 16, both at the same flux level ($\sim0.44$~mJy). We estimated
the 2-10~keV X-ray fluxes of the observations in 1998 and 2002 analysing the
X-ray energy spectrum of the RXTE/PCA observations coordinated to the radio
detection (see also Gilfanov et al. 1998 and Gierlinski, Done \& Barret
2002). The 2-10~keV luminosity of the observation on 2005 June 16 was
estimated using the ASM count rate of $1.9\pm0.4$. We fitted the PCA energy
spectra in the range 3-20~keV with a Gaussian emission line at
$\sim6.4-6.6$~keV, a blackbody with temperatures of kT$\sim0.6-0.7$~keV and
a power-law with a photon index of $\sim1.9$ (the equivalent hydrogen column
density was fixed to N$_{\rm H}=4\times10^{21}$~cm$^{-2}$).

{\em IGR~J00291+5934}: the newly discovered accreting ms X-ray pulsar
IGR~J00291+5934 (Eckert et al. 2004; Markwardt, Swank \& Strohmayer 2004) has
been detected in radio at 15~GHz at the peak of the (soft) X-ray outburst
(Pooley 2004). In Table~1 we show the X-ray flux reported in Markwardt et
al. (2004), radio flux densities in Pooley (2004), and the estimated distances
in Galloway et al. (2005). The radio flux density at 15~GHz has been converted
to 8.5~GHz assuming an optically thin emission with $\alpha=-0.6$.

\begin{figure*}
\begin{tabular}{c}
\psfig{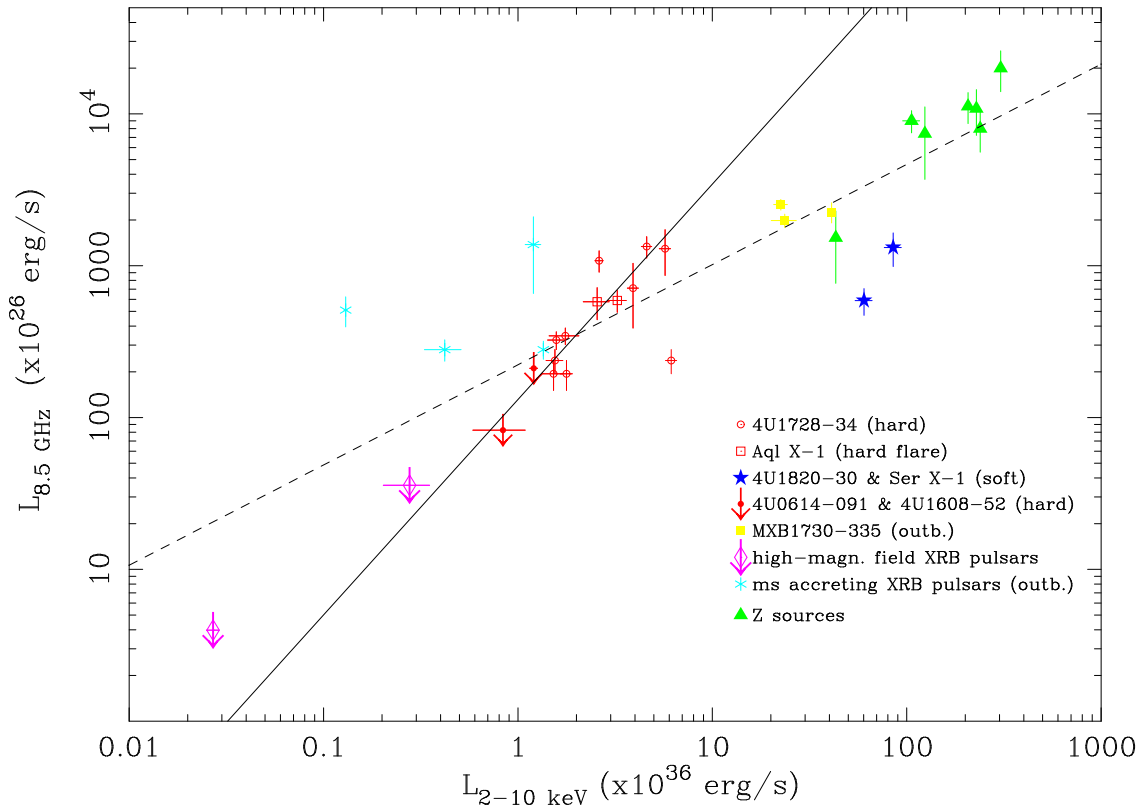}\\
\end{tabular}
\caption{Radio (8.5~GHz) luminosity as a function of X-ray (2-10~keV)
luminosity of NS XRBs: atoll sources in hard state (4U~1728-34: open
circles; Aql~X-1: open squares; 4U 1608-52 and 4U 0614-09: filled
circles with radio upper limits), atoll sources steadily in soft state
(4U~1820-30 and Ser~X-1: filled stars), an `atoll' source in X-ray
outburst (MXB 1730-335: filled squares), accreting ms X-ray pulsars in
X-ray outbursts (SAX J1808.4-3658 and IGR J00291+5934: asterisks; IGR
J00291+5934 is the one with the highest radio luminosity), the
high-magnetic field XRBs (open diamonds with radio upper limits), and
Z sources (filled triangles). The solid line is the fit to the NSs in
hard state, i.e. 4U~1728-34 and Aql~X-1, with a slope of
$\Gamma\sim1.4$ and the dashed line is the fit to all the atolls and Z
sources (excluding only the ms accreting X-ray binaries and the radio
upper limits) with a slope of $\Gamma\sim0.7$ (see
$\S~\ref{section:results_x-r}$).}
\label{NSs} 
\end{figure*}

\subsection{Other high-magnetic field NS XRBs}
{\em 4U 2206+54}: Blay et al. (2005) reported results from INTEGRAL
and VLA observations of the XRB 4U 2206+54. High-energy spectral
analysis reveals the presence of an absorption line at 32~keV, which
indicates the presence of a cyclotron scattering feature, thus
identifying the XRB as a high-magnetic field
($\sim3.6\times10^{12}$~G) NS. VLA observations at 8.5~GHZ did not
detect the radio counterpart with a $3\sigma$ upper limit of
0.039~mJy. In Table~1 we report the distance from Negeruela \& Reig
(2001), the radio flux density in Blay et al. (2005) and the 2-10~keV
X-ray flux from the RXTE/ASM simultaneous daily-averaged count rate.

\subsection{Z-type NS XRBs}

For all the Z sources (Sco~X-1, GX~17+2, GX~349+2, Cyg~X-2, GX~5-1, GX~340+0)
and for GX~13+1 we have calculated the mean X-ray flux based on the ASM count
rate since the beginning of the RXTE mission until Dec 14, 2004.  The mean
radio flux is from Fender \& Hendry (2000). Note that, even though the
classification of GX~13+1 as either atoll- or Z-type NS, is controversial
(e.g. Schnerr et al. 2003), we decided to list it among Z sources because, as
far as radio power is concerned, this source seems to be part of this group
(see Homan et al. 2004).

\subsection{Transient BH XRBs}

In our sample we consider the jet ejection events associated with X-ray
outbursts of eight transient BH XRBs [GRS~1915+105 (flare), GRO~J1655-40,
XTE~J1859+226, XTE~J1550-564, GX~339-4, V4641~Sgr, Cyg~X-1 and XTE~J1748-228]
from Fender et al. (2004 and references therein). We used the values for
distance and radio and X-ray luminosities listed in Table~1 in Fender et
al. 2004; the radio flux densities at 5~GHz has been converted to flux
densities at 8.5~GHz, assuming a radio spectral index $\alpha=-0.6$, and the
fraction of X-ray Eddington luminosity at the peak of the outburst in the
very-high state has been converted to 2-10~keV luminosity, assuming that the
latter is 80\% of the bolometric flux (see \S~\ref{section:edd}).

\subsection{Persistent BH XRBs}

For clarity, we plot in Fig.~2 only GX~339-4 as a sample of BH XRBs in hard
state (for a complete sample of BH XRBs see Gallo et al. 2003). We used fluxes
form Corbel et al. (2003); the 2-10~keV X-ray fluxes have been extrapolated
from the 3-9~keV ones they reported, assuming a spectral index of 1.7. We
calculated the luminosities assuming the (minimum) distance of 7~kpc, inferred
by Zdziarski et al. (2004).

\subsection{Conversion from 2-10~keV luminosities to Eddington units}
\label{section:edd}

In order to extrapolate the bolometric flux of the XRBs, and then to
convert their X-ray luminosities in Eddington units, we have divided
the XRBs in five main groups: BHs in the hard state, BHs during X-ray
outbursts, atoll sources in the hard state, atoll sources in the soft
state, and Z sources. We assumed that each group has the same fraction
of bolometric luminosity in the 2-10~keV band. For each group we used
the best-fit model parameters for the PCA-HEXTE energy spectra to
create a simulated spectrum with {\em xspec}. For the NSs, we used PCA
and HEXTE response matrices and ancillary files to calculate the flux
in the range 3-200~keV, and a Chandra HETGS (MEG) response matrix and
ancillary file to extend the range below 3~keV, down to 0.5~keV,
especially important for soft X-ray states. The 0.5-200~keV has been
taken as a good approximation of the bolometric flux of the
sources. For the BHs in hard state, we used as bolometric fluxes the
3-200~keV fluxes of GX~339-4 in Nowak, Wilms \& Dove (2003). For the
BHs during outbursts we used the bolometric fluxes calculated in
Fender et al. (2004). The conversion to Eddington units is given
dividing the bolometric luminosity by $1.3\times10^{38}\times({\rm
M}/{\rm M}_{\odot})$~erg/s, where M=1.4~M$_{\odot}$ for all the NSs,
while for the BHs we used the masses listed in Table~1 of Fender et
al. (2004).  The bolometric flux will be F$_{2-10}=F_{bol}\times\xi$,
where we used $\xi$=0.2 for BHs in the hard state, $\xi$=0.8 for BHs
during X-ray outbursts, $\xi$=0.4 for atoll sources in the hard state,
$\xi$=0.7 for atoll sources in the soft state and $\xi$=0.8 for Z
sources: the actual $L_{2-10~keV}/L_{bol}$ ratio of the single
observations are always within 10\% of these $\xi$ values.

\section{Results}

\begin{figure*}
\begin{tabular}{c}
\psfig{figure=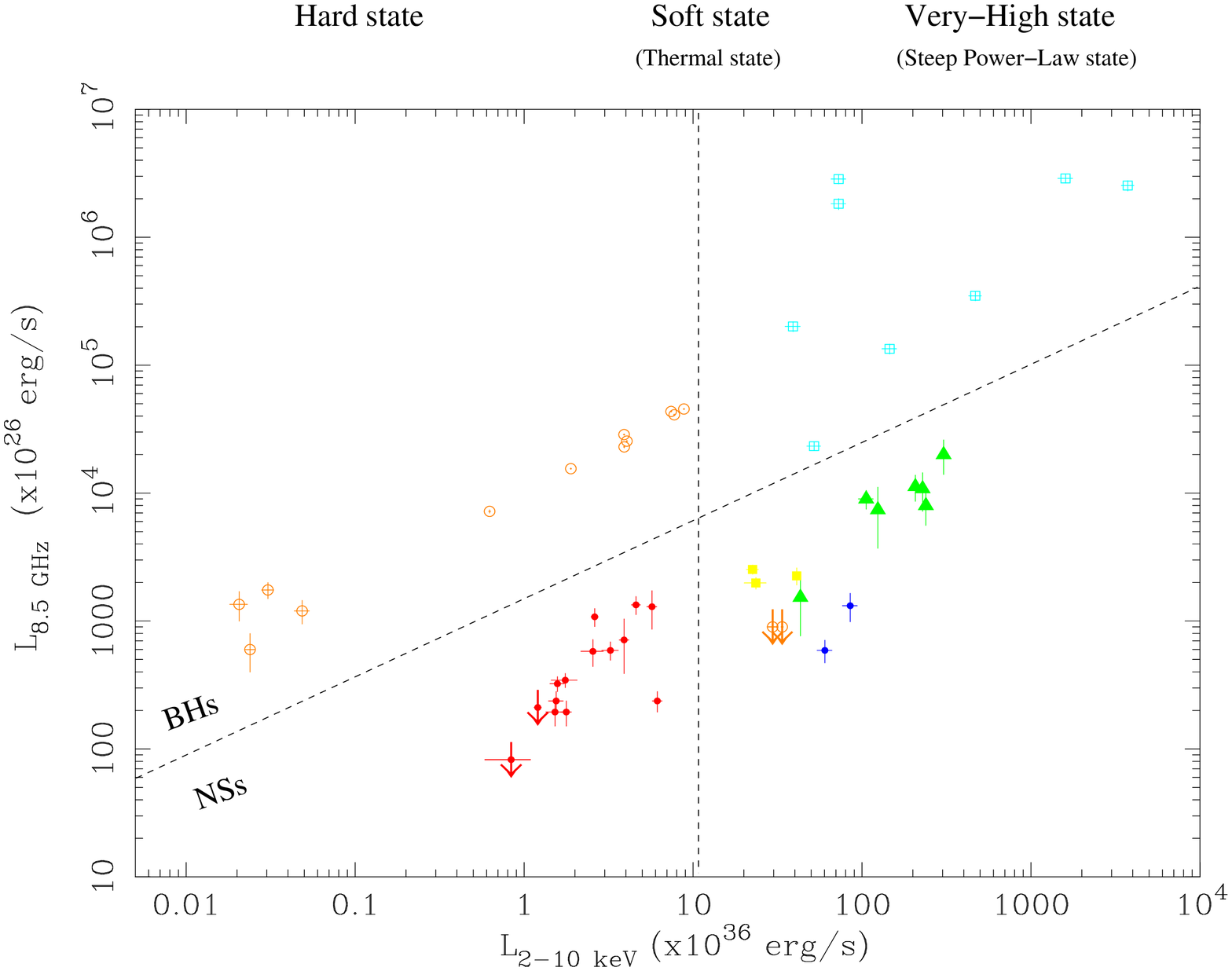,width=15cm,angle=0}\\
\end{tabular}
\caption{Radio (8.5~GHz) luminosity as a function of X-ray
(2-10~keV) luminosity for NS and BH XRBs: GX~339-4 in hard state (open
circles), transient BHs (open squares), atoll sources steady in a hard or soft
state (filled circles), MXB 1730-335 during a soft outburst (filled squares),
and Z sources (filled triangles).} \label{Lx-Lr}
\end{figure*}

\subsection{X-ray/radio luminosities in NS XRBs}
\label{section:results_x-r}

In Fig.~\ref{NSs} we show the radio/X-ray luminosity plane with all
the NS XRBs in our sample. Four groups of sources are plotted:
Z-sources, atoll sources in the hard state, atoll sources steadily in
the soft state and sources in soft outbursts (i.e. the Rapid Burster
and the two accreting ms X-ray pulsars).  There is an overall positive
ranking correlation between radio and X-ray luminosities, at a
significance level of $>99$ per cent. The fit with a power-law to the
Z- and atoll-type NS XRBs (excluding the ms X-ray pulsars) gives a
slope of $\Gamma = 0.66\pm 0.07$ (where L$_{R}\propto
$L$_{X}^{\Gamma}$).

Z-sources (triangles) lie towards the top-right part of the plot, with
X-ray and radio luminosities higher than atolls. We have plotted the
mean of the radio and X-ray luminosities: their radio luminosities are
the superposition of optically thick emission and optically thin
flaring activity, while the X-ray luminosities are the average of the
luminosities in their three possible X-ray states (see a more detailed
discussion in \S~\ref{section:Z-sources}). There is only marginal
evidence for a positive ranking correlation between radio and X-ray
luminosities in the Z sources as a separate group ($\sim96$ per cent
significance level; power-law fit index $\Gamma = 1.08 \pm 0.22$). 

Atoll sources in the hard X-ray state (4U~1728-34: open circles;
Aql~X-1: open stars) show a positive correlation between radio and
X-ray luminosities over one order of magnitude in X-rays (with the
exception of the point with the highest X-ray luminosity: see
discussion in Migliari et al. 2003): a rank-correlation test gives a
significance of $>99$ per cent. In order to compare the luminosity
correlations in NSs with those in BHs (see also
\S~\ref{section:nsvsbh}), we fitted the correlation with a power-law
model: 4U~1728-34 gives $\Gamma=1.40\pm0.25$, and considering also
Aql~X-1 we obtain $\Gamma=1.38\pm0.23$. We should stress once more
that the NS observations span a range of only about one order of
magnitude in X-ray luminosity, to be compared with the three orders of
magnitude of the BH XRBs (see Fig.~\ref{Lx-Lr}). However, we can place
constraints on the slope of the power-law over a larger range of
luminosities. If we consider also the radio upper limits of 4U~0614+09
at low X-ray luminosities, fitting them as detections, we obtain a
lower limit on the slope of the power-law of $\Gamma>1.60\pm0.27$,
clearly indicating that the radio/X-ray luminosity correlation in NSs
is steeper than that in BHs. This has important consequences for our
understanding of the relation of $L_{\rm X}$ and $L_{\rm R}$ to the
accretion rate $\dot{m}$ as shall be discussed later. [Note that the
overall flatter slope of the radio/X-ray luminosity correlation
considering the whole sample of NSs ($\Gamma\sim0.6-0.7$, see above)
is dominated by the slopes within the transients and between the
hard-state sources and the transients.]

\begin{figure*}
\begin{tabular}{c}
\psfig{figure=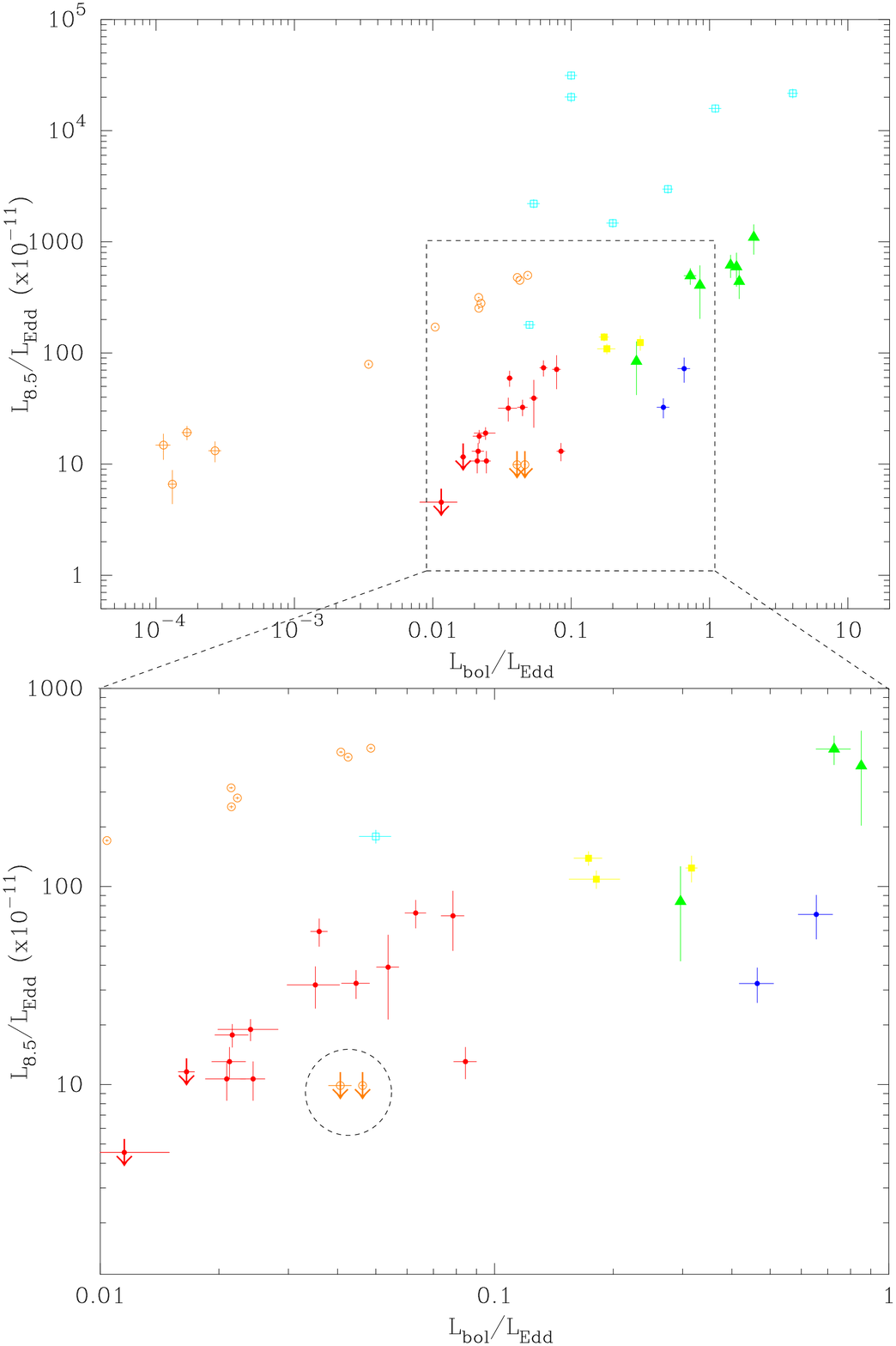,width=13cm,angle=0}\\
\end{tabular}
\caption{ {\em Upper panel:} Same as Fig.~\ref{NSs}, but in Eddington
units (see \S~2.1). {\em lower panel:} Zoom in the range 0.01-1
L$_{Edd}$ showing that in this range there is a radio quenching in BHs
(dashed circle), while in NSs the suppression of the jet is not
observed (at least as extreme as in BHs).}\label{quenching}
\end{figure*}

Atoll sources steadily in soft state (filled stars) have been detected in
radio. This is contrary to what found in BHs, where there is a quenching of
radio emission in the soft state (see Fig.~\ref{quenching}). This finding
indicates that NSs may not suppress completely the (compact?) jet in the soft
state. In fact, considering the ensemble of neutron star data points, there is
no strong evidence {\em at all } for suppressed radio emission in steady soft
states.

The Rapid Burster (filled squares) shows radio flaring emission associated
with X-ray outbursts. It has X-ray luminosities consistent with atoll sources
in the soft state. There is a significant ($99$ per cent) positive ranking
correlation between radio and X-ray luminosities in atoll sources plus the
rapid burster, suggesting that it lies on a sort of natural extension of
atolls in hard state (as in persistent and transient BHs; see Fender et
al. 2004).

The radio peak of IGR J00291+5934 is consistent with the rapid burster radio
peak and with the highest radio emission from 4U~1728-34 (maybe also in a
radio flaring emission state; see Migliari et al. 2003). SAX J1808.4-3658 has
been detected in radio a few days after the peak of the outburst in 1998, when
the X-ray and radio emissions already faded (but see Gaensler et al. 1999) and
during the outbursts in 2002 and 2005 (Rupen et al. 2002, 2005). The radio
luminosities seem to be consistent with those of Aql~X-1, lower than those of
IGR J00291+5934. Additional discussion of the radio emission from the millisec
X-ray pulsars is presented in Migliari, Fender \& van der Klis (2005), in
which it is suggested that they {\em may} be slightly less radio-loud than
other atoll sources as a result of a generally higher surface magnetic field
(Chakrabarty 2005).

The high-magnetic field NSs (X~Per and 4U~2206+54) have not been detected in
the radio band. The radio upper limits are still consistent with the
radio/X-ray luminosity expected extrapolating the correlation for atoll
sources to lower X-ray luminosities. This in fact means that we cannot
confidently state that the high-magnetic field NS are significantly fainter in
the radio band than `normal' atoll sources, when at relatively low ($<10^{-2}$
Eddington) luminosities.

\subsection{Neutron stars  vs. black holes}\label{section:nsvsbh}

Is the `fundamental plane of BH activity' also a fundamental plane for
NSs?  Put in another way, is the X-ray : radio coupling in accreting
black holes related exclusively to the properties of the accretion
flow, or also to some property unique to black holes? Clearly we may
attempt to address this question by comparing the X-ray : radio
coupling in NS XRBs with that of BH in XRBs, and AGN.

Observationally, there are clear qualitative similarities in the disc-jet
coupling between neutron stars and black holes (see Fig.~\ref{Lx-Lr} and
Fig.~\ref{quenching}):

\begin{itemize}
\item{below a certain X-ray luminosity, in hard
X-ray states (i.e. $L_{\rm X} < 0.1\times L_{\rm Edd}$), both classes
of objects seem to make steady, self-absorbed jets (caveat very poor
measurements of radio spectra in the case of NSs) which show correlations 
between $L_{X}$ and L$_R$.}
\item{at higher X-ray luminosities, close to the Eddington limit,
bright, optically thin, transient events occur (specifically
associated with rapid state changes).}
\end{itemize}

These similarities indicate that the coupling between the jet and the
innermost regions of the accretion disc does not depend (at least
entirely) on the nature of the compact object, but it is related to
the fundamental processes of accretion in strong gravity.

However, there are quantitative differences in the disc-jet coupling also:

\begin{itemize}
\item{The neutron stars in the hard state appear to show a steeper dependence
of $L_R$ on $L_X$, also with a lower normalisation in $L_R$.}
\item{The neutron stars do not appear to show anywhere near as much
suppression of radio emission in steady soft states as the black
holes}
\end{itemize}

We performed a Kolmogorov-Smirnov test on the ratios between L$_{X}$ and
L$_{R}$ in the two XRB systems, to check if the BHs and NSs X-ray/radio
luminosities are drawn from the same distribution. The null hypothesis that the
data sets are drawn from the same distribution is $\sim10^{-3}$ for the
observations in the hard state only (i.e. GX~339-4 vs. 4U~1728-34 and Aql~X-1)
and $\sim10^{-5}$ using the whole sample. This indicates clearly a different
dependence of L$_{R}$ over L$_{X}$ in the two systems.

In fact, whether in absolute units, Eddington-scaled units, or
applying the mass-correction appropriate for the 'fundamental plane of
black hole activity' ($L_R \propto M^{0.8}$; Merloni et al. 2004), the
neutron stars remain stubbornly less radio-loud than the black holes
for a given X-ray luminosity. Bolometric corrections are however only
poorly estimated at lower luminosities, and could conceivably bring
the data sets significantly closer together if severely underestimated
for the BH sample.

\section{Discussion}

In the following we will briefly discuss some possible implications deriving
from the comparison between disc-jet coupling in BHs and NS systems.

\begin{figure}
\begin{tabular}{c}
\psfig{figure=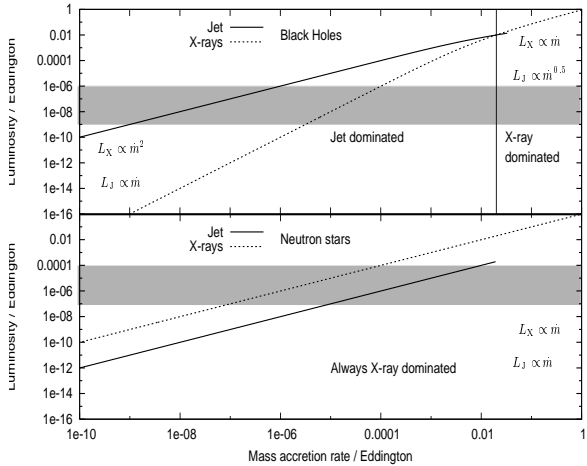,width=8cm,height=7cm,angle=0}\\
\end{tabular}
\caption{Computed variation of jet and X-ray power as a function of
mass accretion rate (all in Eddington units) for black holes (upper
panel) and neutron stars (lower panel). This is a reworking of the
model presented in Fender, Gallo \& Jonker (2003), based upon the new
observational evidence presented in this paper and elsewhere. We now
presume that in all hard X-ray states black holes are jet-dominated,
whereas neutron star systems {\em never} enter the jet-dominated
regime. The grey bars indicate the observed 'quiescent' X-ray
luminosities of most X-ray transients.  }\label{jetdom}
\end{figure}

\subsection{Jet velocity and power}

Observations of the ultrarelativistic radio jets in the NS XRB Cir~X-1
(i.e. with a bulk Lorentz factor $>15$; Fender et al. 2004) have
already shown that the (often accepted) `escape velocity' paradigm,
which states that the jets' velocity should be about the escape
velocity of the compact object involved, is not valid in the
relativistic regime. Their observations also indicate that properties
unique to BHs are {\em not necessary} for the production of
relativistic jets. However, characteristics proper of the compact
object seem to play, at least partially, a role in the jet production.

Regarding radio and jet power, it is important to know at what $L_{\rm X}$ to
compare the NS and BH samples. We would argue (see below) that the least
radiatively inefficient point, while still in the hard state and therefore
producing a steady jet in both samples, should be selected. This point
naturally corresponds to the brightest low/hard / hard-atoll
states. Comparison of the fits to the NS and BH samples indicates that at
$L_{\rm X} \sim 0.02$, the ratio of radio luminosities is $\sim 30$. As noted
in Migliari et al. (2004), assuming a scaling L$_{R} \propto
L_{J}^{1.4}$ this indicates that neutron star jets are about one order of
magnitude less powerful than black hole jets at this X-ray luminosity. As we
shall see below, the diverging L$_{R}$:$L_{\rm X}$ correlations do
{\em not} require that this ratio change as a function of accretion rate.

\subsection{Event horizons and radiatively inefficient flows}

The different correlations between $L_{\rm X}$ and L$_{R}$ in the BH
and NS samples (to recall, $b_{BH} \sim 0.7$, $b_{NS} \ga 1.4$, where $L_{\rm
radio} \propto L_{X}^b$) are telling us something quite fundamental about the
accretion processes in these two types of object. In the following we shall
take $b_{NS} = 1.4$. Assuming, as before, that L$_{R} \propto
L_{J}^{1.4}$, we get

\begin{tabular}{cc}
BH & $L_J \propto L_X^{0.5}$ \\
NS & $L_J \propto L_X$ \\
\end{tabular}

where the quadratic relation for the black holes was already presented in
Fender, Gallo \& Jonker (2003). The linear relation between jet and X-ray
powers in the NS sample implies that neutron star systems will never reach a
jet-dominated state (unless sources like 4U 1728-34 are already in
jet-dominated states, but this seems unlikely).

The different relations may seem to imply, at face value, that the coupling
between accretion rate and jet power may be different in these two sets of
sources. If we assume that the relation between $L_X$ and $\dot{m}$ is the
same for both BH and NS this is clearly true. However, we believe it is far
more likely that it is the coupling between $L_X$ and $\dot{m}$ which is
different in the two samples, as we shall outline below.

Assuming that the relation between $L_J$, and not $L_X$, and the accretion
rate $\dot{m}$ is the same for both classes of object, we can draw some simple
yet important conclusions. Assuming that accretion in NS sources is
essentially radiatively efficient (in the presence of a solid surface, the
only way to avoid this criterion is if a large fraction of the accreting {\em
mass} were ejected before it had radiated or impacted on the NS surface), then
for NS we get simple linear relations:

\begin{tabular}{cc}
NS & $L_J \propto L_X \propto \dot{m}$\\
\end{tabular}

and since $\dot{m}\propto L_X+L_J$ and we estimate $L_X > L_J$, then
in Eddington units

\begin{tabular}{cc}
NS & $L_X \sim \dot{m}$ \\
\end{tabular}

Keeping the same coupling between accretion rate and jet
power for black holes, we arrive at

\begin{tabular}{cc}
BH & $L_J \propto L_X^{0.5} \propto \dot{m}$ \\
\end{tabular}

This is exactly the prescription presented in Fender, Gallo \& Jonker
(2003) for jet-dominated states in X-ray binary systems. Therefore,
one clear explanation for the observed differences between the
L$_{R}$:$L_X$ correlations in the two samples is that the NS are in a
'X-ray dominated' state and the BH are 'jet-dominated'. The different
coupling between $L_X$ and $\dot{m}$ ensures that the samples remain
fixed in these states as the accretion rate decreases. In
Fig.~\ref{jetdom} we plot the situation as we now envisage it. Note
that we have adopted jet power normalisations of $A_{\rm steady, BHC}
= 0.1$, $A_{\rm steady, NS} = 0.01$ (where
$L_{J}=A\times L_{R}^{1/1.4}$; Fender, Gallo \& Jonker 2003; Fender,
Maccarone \& van Kesteren 2005). The value of $A_{\rm steady, BHC} =
0.1$ corresponds to equipartition between jet and X-ray powers at
around the soft $\rightarrow$ transition luminosity of $L_{\rm X,
 trans} \sim 0.02$. This is a larger normalisation than the
conservative lower limit presented in Fender, Gallo \& Jonker (2003)
but we consider it to be more likely given the lack of apparent
accretion efficiency transitions {\em within} the hard state (see
discussions in e.g. Malzac, Merloni \& Fabian 2004; Maccarone 2005)
and recent, higher, estimates of the steady jet power (e.g. Gallo et
al. 2005).  In this framework, the difference in quiescent
luminosities of BH and NS X-ray binaries (e.g. Garcia et al. 2001) are
simply explained by the jet removing most of the liberated
gravitational potential energy in the quiescent BH, but not in the NS,
conclusions identical to those drawn in Fender, Gallo \& Jonker
(2003). Accretion rates $10^{-6} \la \dot{m} \la 10^{-4}$ (Eddington
units) for both classes of object in quiescence can produce the
observed discrepancy in $L_X$ (Fig.~\ref{jetdom}).

However, the result that for the BH $L_{X} \propto \dot{m}^2$ is generically
indicative of {\em radiatively inefficient} accretion in the black hole
systems. We define radiatively inefficient to mean that the majority of the
liberated gravitational potential is carried in the flow and not radiated
locally; in this sense the jet-dominated configuration outlined above
corresponds to radiatively inefficient accretion, since most of the liberated
accretion power is in the form of the internal and bulk kinetic energy of the
ejected matter. There is of course another appealing possibility, namely that
we are witnessing the observational effect of advection-dominated accretion
flows (ADAFs, e.g. Ichimaru 1977; Narayan \& Yi 1994, 1995), in which case the
discrepancy between $L_X$ and $\dot{m}$ corresponds to the majority of the
liberated gravitational potential energy being advected across the black hole
event horizon.

Clearly, despite their similarities in being radiatively inefficient accretion
configurations, the jet-dominated scenario and the ADAF model are very
different. Estimates of the jet power normalisation (see above) indicate that,
in our opinion, ADAF-like solutions in which most of the available
gravitational potential crosses the event horizon are not {\em required} by
the observations. However, uncertainties in the estimates of the jet power
normalisation, the true accretion rate, etc.  mean that it may still be an
important, even dominant, channel. Models of radiatively inefficient accretion
flows in which powerful outflows are driven (e.g. Blandford \& Begelman 1999)
may be the most appropriate. It is worth noting that relation $L_J \propto
\dot{m}$ is similar / identical to several previous models of jet powering
(e.g. Falcke \& Biermann 1996; Meier 2001).

\subsection{The role of the magnetic field}

It is a general accepted idea that very high-magnetic fields at the
surface of the NSs inhibit the production of {\em steady} jets (while
a large amount of energy can be extracted from magnetic fields to
power extremely energetic transient jets, as e.g. in the case of the
magnetar SGR~1806-20; Gaensler et al. 2005; Cameron et
al. 2005). However, besides theoretical arguments, actual
observational proves are missing. The upper limits on previous
observations (e.g. Fender \& Hendry 2000 and references therein),
although significantly lower than radio detections of BH XRBs, are not
at all stringent if compared with other NS sources detected in radio,
and actually higher than the radio detection levels of atoll sources
at the same accretion rate (as traced by the X-ray
luminosity). Chakrabarty (2005) suggested that accreting ms X-ray
pulsars have a slightly higher magnetic field than other atoll
sources. This would suggest that we should see a decreasing radio
luminosity (for a given mass accretion rate) from atoll sources to
accreting ms X-ray pulsars to high-magnetic field X-ray pulsars. Note
that all the radio detections of the accreting ms X-ray pulsars have
been made during outburst and no information is available of their
steady compact jet, whose radio power should be anyway lower than the
transient jet detections (see also discussion about ms X-ray pulsars
in Migliari et al. 2005).  Although high-magnetic field NS XRBs have
not yet been detected in the radio band, their upper limits (the
lowest upper limits to date are shown in Fig.~\ref{NSs}), are still
consistent with the extrapolation at low X-ray luminosities of the
radio/X-ray luminosity correlation of the low-magnetic field NS
XRBs. Up-coming radio observations of high-magnetic field and
accreting ms X-ray pulsars will give us the opportunity to test these
ideas, and quantify the role of the magnetic field in the jets
production.

\subsection{X-ray timing features and radio jet power}

There is a correlation between the radio luminosity and the characteristic
frequencies of the low-frequency timing components in the X-ray power spectra
in NS and BH XRBs (Migliari et al. 2005): the timing features are direct
tracers of the radio jet power. The fitting power-laws of the correlations
between radio luminosity and the characteristic frequencies of the L$_{h}$
Lorentzian component of the power spectrum in NSs and the L$_{\ell}$
Lorentzian component in the BH GX~339-4 are:
L$_{R}\propto\nu_{h}^{1.30\pm0.10}$ and
L$_{R}\propto\nu_{\ell}^{1.37\pm0.02}$.

Timing features are related to accretion disc properties, and in
particular kHz quasi-periodic oscillation (QPO) frequencies are
generally interpreted as being related to the motion of matter in the
accretion disc at a preferential radius, very close to the compact
object (see van der Klis 2004 for a review).  In XRBs, all the
variability components in the power spectra follow a universal scheme,
when plotted against the upper-kHz QPO (e.g. Psaltis, Belloni \& van
der Klis 1999; Belloni, Psaltis \& van der Klis 2002; van Straaten et
al. 2002; van Straaten, van der Klis \& M\'endez 2003; Altamirano et
al. 2005; Linares et al. 2005), therefore $\dot{m}$ may be in
principle inferred also by low-frequency timing features. In
particular, a tight correlation exists between the characteristic
frequency of the upper-kHz QPO $\nu_{u}$ and $\nu_{h}$ in atoll
sources: the best-fit power-law is
$\nu_{h}\propto\nu_{u}^{2.43\pm0.03}$ (van Straaten, van der Klis \&
Wijnands 2005). In jet models, the total power of a steady compact
jets is related to the radio power as L$_{R}\propto$L$_{J}^{1.4}$
(Blandford \& Konigl 1979; Falcke \& Biermann 1996; Markoff, Falcke \&
Fender 2001)). A linear relation between L$_{J}$ and the mass
accretion rate $\dot{m}$ is suggested by the comparative quantitative
study of the radio/X-ray luminosity correlations between NS and BH
XRBs in hard X-ray state (see above). If this scaling is correct,
$\nu_{h}$ in NSs and $\nu_{\ell}$ in BHs scale about linearly with
$\dot{m}$. Using the $\nu_{h}\propto\nu_{u}^{2.43\pm0.03}$ empirical
correlation and L$_{R}\propto\nu_{h}^{1.30\pm0.10}$ found in atoll
NSs, we obtain a relation that link about quadratically the upper kHz
QPO frequency (possible indicator of the inner disc radius) and the
mass accretion rate: $\dot{m}\propto\nu_{u}^{2.16\pm0.20}$.

Furthermore, these relations can be important for a direct comparison
to AGN. In particular, the relation between the radio luminosity and
the `break' frequency (Migliari et al. 2005), a timing feature
observed also in AGN (e.g. McHardy et al. 2005), opens the possibility
of the existence of a 'new fundamental plane' for BHs. Taking into
account the mass scaling, we can directly compare stellar-mass and
supermassive black holes in a three-dimentional space where the
variables are the mass of the black hole, the radio luminosity and the
frequency of the break component (which is independent from the
distance to the source). The existence of another `fundamental plane'
would further support the idea of a unified description of the
coupling between disk and jet in black holes of all masses, and
possibly including also NSs.

\subsection{Z sources: the NS equivalents of GRS 1915+105}\label{section:Z-sources} 

\begin{figure}
\begin{tabular}{c}
\psfig{figure=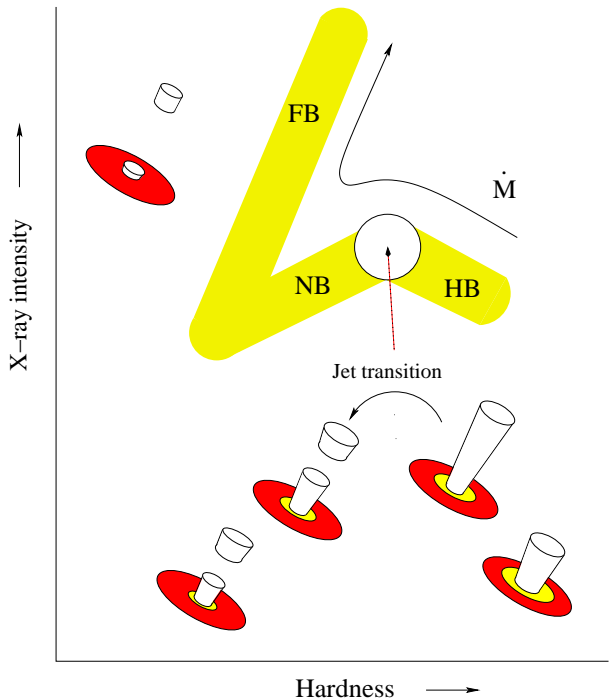,width=8cm,angle=0}\\
\end{tabular}
\caption{Sketch of the disc-jet coupling in Z-type sources, which we believe
to be the neutron star equivalents of the black hole GRS 1915+105, constantly
accreting at $\sim$Eddington rates and producing powerful jets associated with
rapid state transitions. See \S~\ref{section:Z-sources} for a discussion}\label{Z_HID} 
\end{figure}

Studying the disc-jet connections in BH XRBs during X-ray outburst events
(throughout transitions between X-ray states), Fender et al. (2004) developed
a sketch-model which shows how the accretion disc properties (as traced in
X-rays) are connected to the jets production (traced with radio). In their
picture, sources like GRS~1915+105 are persistently at the `edge' between the
X-ray state in which the source produces a core compact jet (hard state), and
the state [very-high state or steep power-law state in the nomenclature
introduced by McClintock \& Remillard (2005)] in which the compact jet is
disrupted and a radio optically thin flare (associated to a fast ejection of
matter) is observed. The semicontinuous crossing of the `line' between the
hard and the soft states could explain the sequence of rapid radio flares
observed in this source (Fender \& Belloni 2005; Fender et
al. 2005). Z-type NS XRBs, strongly variable in X-rays, show rapid variability
also in the radio band, where we observe besides the approximately steady
optically thick emission, frequent optically thin flares (see Fig. 6). These
optically thin radio flares are possible signatures of fast ejected plasmons,
already observed as extended lobes in Sco~X-1 (Fomalont et al. 2001a,b) and
Cir~X-1 (Fender et al. 1999). Z sources, which are always near or at the
Eddington accretion rate, seem to be like e.g. GRS~1915+105 semicontinuously
at the edge between the state in which a powerful core compact jet still
exists and the launch of optically thin plasmons which follow the disruption
of this compact jet. We might say that Z sources are the NSs `counterpart' of
transient BHs at very high accretion rates.

\subsubsection{A disc-jet coupling for Z-type neutron stars}

In four Z-type NSs (Sco X-1, GX~17+2, Cyg X-2 and GX~5+1) an association
between the position on the CD and the radio flux have been reported, the
radio flux decreasing from HB to NB to FB (Penninx et al. 1988; Hjellming et
al. 1990a,b; with the exception of GX~5+1 for which the radio flux is higher
in the NB than in the HB; Tan et al. 1992). Z sources show a variable radio
activity, where rapid and powerful flares are often observed besides a more
steady radio emission. Sco X-1, in particular, is the only Z source among the
four for which the extended radio jets have been spatially resolved moving
away form the radio core. Sco X-1 has been observed for 56 hr on June 11-13,
1999 (MJD~$51340-51342.5$) simultaneously in radio with the Very Long Baseline
Interferometer (VLBI) and in X-rays with RXTE. The results are reported in
Fomalont et al. (2001b; radio analysis) and Bradshaw, Geldzahler \& Fomalont
(2003; X-ray analysis). These are the most complete observations of a Z source
we have to date in order to study the disc-jet coupling. The radio activity of
Sco X-1, i.e. the flux and spectral evolution of each of the spatially
resolved radio components (core, north-west lobe and south-east lobe), can be
monitored in relation to the changes of the X-rays properties (e.g. position
on the CD). In the following we will concentrate on these observations, in
particular using Fig.~3 and Fig.~4 in Fomalont et al. (2001b) and Table~1 and
Fig.~1 in Bradshaw et al. (2003), and will attempt to draw a phenomenological
disc-jet coupling model accounting for these and the other Z sources'
observations (see Fig.~\ref{Z_HID}).

We follow in detail the evolution of two radio components of Sco X-1, the core
and the north-west (NW) extended jet, from MJD~$51340$ to MJD~$51342.5$
(Fig.~3c,d and Fig.~4a,b in Fomalont et al. 2001). From Table~1 and Fig.~1 of
Bradshaw et al. (2003) we know that Sco X-1 is mainly in the HB on
MJD~$51340$, in the NB on MJD~$51341$ and in the FB on MJD~$51342$.

\begin{itemize}

\item On MJD 51340 (HB) the radio flux of the core rises. The radio spectrum is
optically thick, indicating that the radio emission likely comes from a
compact jet. Contemporaneously the NW extended jet is fading, meaning that it
is still decoupled from the activity of the core jet.

\item On MJD 51340 (NB) the core shows optically thin radio emission,
suggesting a renewed transient ejection activity (not yet spatially
resolved). Note that Fomalont et al. (2001a) already noted that flares in the
core are followed by flares in the NW lobe indicating (unseen) relativistic
ejections from the core. Put in another way, what we see in the core is likely
the superposition of the optically-thick compact jet and of the optically-thin
emission from discrete plasmon ejections. Making a parallel with the behaviour
of BHs where transient jets are associated with X-ray state changes
(e.g. Mirabel et al. 1999; Gallo et al. 2003; see Fender et al. 2004), we can
associate the HB-to-NB state change with the ejection of transient jets [in
BHs the transient jets are associated with the VHS (or Steep Power-law state);
Fender et al. 2004]. Around MJD 51341.5, the flux in the core decreases while
the source is in the FB, to increase again in correspondence of the FB-to-NB
transition.  We do not have a dual-frequency monitoring during this period, so
we cannot know the nature of the radio spectrum of the flare, although, given
the optically thin decay of the flare observable on MJD 51342, a transient jet
activity associated with the FB-to-NB state change as well is a plausible
scenario.

\item On MJD 51342 (FB) we observe a decay in the core radio flux with an
optically thin radio spectrum. In general, during the FB the source has been
observed to have the lowest radio flux, therefore suggesting a suppression of
the (compact) jet and of the transient plasmons ejection activity (the faint
optically thin emission we observe is possibly the `relict' of a transient jet
previously ejected).

\end{itemize}

In Fig.~\ref{Z_HID} we show the schematic of the disc-jet coupling in Z
sources. The typical HID of a Z source is sketched as a `snake' track. The
mass accretion rate $\dot{m}$ is thought to increase along the track from HB
to FB. Starting from HB, as $\dot{m}$ increases so does the compact jet power.
Crossing the HB-to-NB state transition point (circle on the HID track) a
transient jet is launched. Meanwhile, the compact jet power decreases. When
the source is in the FB the jet activity is quenched, possibly due to a very
high mass accretion rate. The cycle HB-NB-FB-NB-HB lasts no more than a few
days. Therefore, Z sources, like GRS~1915+105, are continuously crossing the
`jet transition' point, showing a frequent transient jet activity.

All the other coordinated X-ray/radio observations of Z sources (Penninx et
al. 1988; Hjellming et al. 1990a,b; Tan et al. 1992) are consistent with this
model. Note that Tan et al. (1992) reported that the radio flux in GX~5+1 is
weaker in the HB than in the NB (contrary to the more simple qualitative
association between an X-ray state and a radio flux: the radio flux decreasing
from the HB to the NB to the FB). Looking at their Fig.~1, we can see that
during the observations on September 1, 1989, the source was in a `very hard'
state, i.e. at the bottom right of the HID track in our Fig.~\ref{Z_HID} (with
the lowest $\dot{m}$), where the compact jet was possibly still not very
powerful. On September 4, 1989, they observed a powerful radio flare when the
source was in the NB where, indeed, we expect (optically thin) flaring
activity.

\section{Conclusions}

Comparing the connections between X-ray and radio properties in NS and BH
systems, we have found many similarities and differences, that can be read in
terms of physical ingredients for the production of jets.\\

{\em i)} Below a certain X-ray luminosity, in hard state (i.e. $L_{\rm
X} \leq 0.02 L_{\rm Edd}$), both classes of objects seem to make steady,
self-absorbed jets while at higher X-ray luminosities, close to the Eddington
limit, bright, optically thin, transient events occur (specifically associated
with rapid state changes).\\

{\em ii)} In the hard X-ray states, correlations between radio and X-rays
emission have been found in both BHs and NSs. This indicates that the link
between the power of the jet and the innermost regions of the accretion disk
does not depend (at least entirely) on the nature of the compact object, but
it is related to the fundamental processes of accretion in strong gravity, and
can be inferred as the mass accretion rate.\\ 
 
{\em iii)} Neutron star X-ray binaries are definitely less 'radio
loud' than black hole X-ray binaries.  At a given X-ray luminosity,
and at a given fraction of Eddington luminosity the BHs produce more
powerful jets than NSs. The difference in radio power is $\ga30$,
which can be reduced to $\ga7$ if we consider possible mass
corrections as derived from the black holes' fundamental plane or to a
factor of $\ga5$ if we consider the mass correction coming from the
conversion of the 2-10~keV luminosities in Eddington units.\\

{\em iv)} Contrary to BHs, atoll-type NSs have been detected in radio
when steadily in soft X-ray states, suggesting that quenching of jet
formation in disc-dominated states may not be so extreme, or that
neutron stars have another channel for producing radio emission.  \\

{\em v)} The slope of the power-law correlation in the hard state of BHs is
$\sim0.7$, while for NSs is steeper (possibly $\ga1.4$)\\

{\em vi)} A power-law slope greater than 1.4 in NSs implies that NSs never
enter a jet-dominated state.\\

{\em vii)} Both the jet-dominated and ADAF frameworks can naturally
explain the difference in slope of the radio/X-ray luminosity
correlations between NSs and BHs, if the total jet power is about
linearly proportional to the disc mass accretion rate:
L$_{J}\propto\dot{m}$. In particular both frameworks derive the same
relations between the X-ray luminosity and the mass accretion rate:
L$_{X}\propto\dot{m}$ for NSs and L$_{X}\propto\dot{m^{2}}$ for BHs.
This is strong independent evidence that the X-rays in hard state
black holes originate in a radiatively inefficient flow, independent
of whether the 'missing' energy escapes to infinity in an outflow or
crosses a black hole event horizon.  \\

{\em viii)} There are correlations between radio luminosity and the
characteristic frequency of X-ray timing components in NSs and in BHs: timing
features are direct tracers of the radio jet power. Assuming a linear relation
between the total jet power and the mass accretion rate, a relation between the
characteristic frequency of the upper kHz QPO and the mass accretion rate can
be inferred: $\dot{m}\propto\nu_{u}^{\sim2}$.\\

{\em ix)} The role in the production of jets of the magnetic field at the
surface of the NS is not clear yet, although it is believed that the higher the
magnetic field the lower should be the jet power: further radio observations
of X-ray pulsars and millisec accreting X-ray pulsars are needed to give
observational constraints and quantify its role. \\

{\em x)} Z-type NSs, which are always near or at the Eddington accretion
rate, seem to be like GRS~1915+105 semicontinuously at the edge between the
state in which a powerful core compact jet still exists and the launch of
optically thin plasmons. Following, in particular, detailed simultaneous
radio/X-rays observations of Sco~X-1 we draw a model that can describe the
disc-jet coupling in Z sources, finding a possible association between the
HB-to-NB state change and the emission of transient jets.

\section*{Acknowledgements}
The Australia Telescope is funded by the Commonwealth of Australia for
operation as a National Facility managed by CSIRO. The Westerbork
Synthesis Radio Telescope is operated by ASTRON (Netherlands
Foundation for Research in Astronomy) with support from the
Netherlands Foundation for Scientific Research NWO. We would like to
thank Elena Gallo for her help in the reduction of the 21 cm WRST data
and for comments on a draft of this manuscript and also Peter Jonker
and John Tomsick for comments on the paper.

\end{document}